\documentclass[seceq]{ptptex}
\usepackage{wrapft}
\usepackage{bm}

\title{
Sfermion Mass Relations in Orbifold Family Unification
}

\author{
Yoshiharu \textsc{Kawamura}\footnote{E-mail: haru@azusa.shinshu-u.ac.jp} 
and Teppei \textsc{Kinami}\footnote{E-mail: s06t303@shinshu-u.ac.jp}
}

\inst{
Department of Physics, Shinshu University, Matsumoto 390-8621, Japan
}

\recdate{
September 11, 2007}

\abst{
We derive relations among sfermion masses based on orbifold family unification models.
Sfermion mass relations are specific to each model and can be useful for a selection of realistic model.
}
\begin{document}

\maketitle

\section{Introduction}

Supersymmetric grand unified theories (SUSY GUTs) on an orbifold have attractive features 
as a realistic model beyond the minimal supersymmetric standard model (MSSM).
The triplet-doublet splitting of Higgs multiplets is elegantly realized 
in the framework of SUSY $SU(5)$ GUT in five dimensions.\cite{K}\cite{H&N}
Four-dimensional chiral fermions are generated through the dimensional reduction.
Those phenomena originates from the fact that a part of zero modes are projected out 
by orbifolding, i.e., by non-trivial boundary conditions (BCs) concerning extra dimensions on bulk fields.
There is a possibility that a (complete) family unification is realized by eliminating 
all mirror particles from the low-energy spectrum.
Mirror particles are particles with opposite quantum numbers under the standard model (SM) gauge group $G_{SM}$.

Recently, the family unification has been studied in SUSY $SU(N)$ GUTs defined on 
a five-dimensional space-time $M^4 \times (S^1/Z_2)$.\cite{KK&O}.\footnote{
The possibility that one might achieve the complete family unification 
utilizing an orbifold has been also suggested in the earlier reference~\cite{BB&K} in a different context.
In Ref.~\citen{Watari:2002tf}, three families have been derived from a combination of a bulk gauge multiplet 
and a few brane fields.
In Ref.~\citen{Chaichian:2001fs}, they have been realized as composite fields.}

Here $M^4$ is the four-dimensional Minkowski space-time and $S^1/Z_2$ is the one-dimensional orbifold. 
A great variety of models have been found, in which zero modes from a single bulk field 
and a few brane fields compose three families, and we refer them as orbifold family unification models.
At present, it is important to make powerful predictions in order to specify models by using experimental data.
Much works concerning mass relations among scalar particles have been carried out 
based on the motivation that relations specific to each model will give a hint to understand the structure of 
the MSSM and beyond in four-dimensional SUSY models\cite{FHK&N,M&R,KM&Y,P&P,K&T,Ch&H,K&M,A&P,IK&Y}.\footnote{
Scalar mass relations have been examined in four-dimensional superstring models.\cite{KK&K,K&K}}
Sum rules among sfermion masses have been also derived in two kinds of orbifold family unification models,
and it has been pointed out that they can be useful probes of each model\cite{K&Ki}.\footnote{
Sfermion masses have been studied from the viewpoint of flavor symmetry and its violation
in $SU(5)$ SUSY orbifold GUT.\cite{H&N2}}

In this paper, we study sfermion masses from a general framework, based on orbifold family unification models 
under some assumptions regarding the breakdown of SUSY and gauge symmetries, and derive relations among them.

This paper is organized as following. 
In \S 2, we explain the outline of orbifold family unification models.
In \S 3, we give a generic mass formula for sfermions and derive specific relations 
among sfermion masses.
\S 4 is devoted to conclusions and discussions.

\section{Orbifold family unification}

First we review the arguments in Ref.\citen{KK&O}.
We study $SU(N)$ gauge theory on $M^4 \times (S^1/Z_2)$ with the gauge symmetry breaking pattern, 
$SU(N) \to SU(3) \times SU(2) \times SU(r) \times SU(s) \times U(1)^n$,
which is realized by the $Z_2$ parity assignment
\begin{eqnarray}
&~& P_0 = \mbox{diag}(+1, +1, +1, +1, +1, -1, \dots, -1, -1, \dots, -1) , 
\label{P0-SM} \\
&~& P_1 = \mbox{diag}(+1, +1, +1, -1, -1, \underbrace{+1, \dots, +1}_{r}, \underbrace{-1, \dots, -1}_{s}) ,
\label{P1-SM}
\end{eqnarray}
where $s = N-5-r$ and $N \ge 6$.
The $n$ is an integer which depends on the breaking pattern.
The matrices $P_0$ and $P_1$ stand for the representation matrices (up to sign factors) of the fundamental representation 
for the $Z_2$ transformation ($y \to -y$) and the $Z_2'$ transformation ($y \to 2\pi R- y$), respectively.
Here $y$ is a coordinate of $S^1/Z_2$ and $R$ is a radius of $S^1$.
After the breakdown of $SU(N)$, the rank $k$ totally antisymmetric tensor representation $[N, k]$, 
whose dimension is ${}_{N}C_{k}$, is decomposed into
a sum of multiplets of the subgroup $SU(3) \times SU(2) \times SU(r) \times SU(s)$
\begin{eqnarray}
[N, k] = \sum_{l_1 =0}^{k} \sum_{l_2 = 0}^{k-l_1} \sum_{l_3 = 0}^{k-l_1-l_2}  
\left({}_{3}C_{l_1}, {}_{2}C_{l_2}, {}_{r}C_{l_3}, {}_{s}C_{l_4}\right) ,
\label{Nk}
\end{eqnarray}
where $l_1$, $l_2$ and $l_3$ are intergers, $l_4=k-l_1-l_2-l_3$
and our notation is that ${}_{n}C_{l}= 0$ for $l > n$ and $l < 0$.
Here and hereafter we use ${}_{n}C_{l}$ instead of $[n, l]$ in many cases.
We list $U(1)$ charges for representations of subgroups in Table \ref{t1}.
\begin{table}
\caption{The $U(1)$ charges for representations of fermions.}
\label{t1}
\begin{center}
\begin{tabular}{c|c|c|c|c} \hline
species & representation  & $U(1)_1$ & $U(1)_2$ & $U(1)_3$ \\ \hline\hline
$(\nu_{R})^c$, $\hat{\nu}_{R}$ & $\left({}_{3}C_{0}, {}_{2}C_{0}, {}_{r}C_{l_3}, {}_{s}C_{k-l_3}\right)$ & $0$ & $(N-5)l_3 - rk$ & $-5k$ \\ \hline
$(d'_{R})^c$, $d_{R}$ & $\left({}_{3}C_{1}, {}_{2}C_{0}, {}_{r}C_{l_3}, {}_{s}C_{k-l_3-1}\right)$ & $-2$ & $(N-5)l_3 - r(k-1)$ & $N-5k$ \\
$l'_{L}$, $(l_{L})^c$ & $\left({}_{3}C_{0}, {}_{2}C_{1}, {}_{r}C_{l_3}, {}_{s}C_{k-l_3-1}\right)$ & $3$ & $(N-5)l_3 - r(k-1)$ & $N-5k$ \\ \hline
$(u_{R})^c$, $u'_{R}$ & $\left({}_{3}C_{2}, {}_{2}C_{0}, {}_{r}C_{l_3}, {}_{s}C_{k-l_3-2}\right)$ & $-4$ & $(N-5)l_3 - r(k-2)$ & $2N-5k$ \\
$(e_{R})^c$, $e'_{R}$ & $\left({}_{3}C_{0}, {}_{2}C_{2}, {}_{r}C_{l_3}, {}_{s}C_{k-l_3-2}\right)$ & $6$ & $(N-5)l_3 - r(k-2)$ & $2N-5k$ \\
$q_{L}$, $(q'_{L})^c$ & $\left({}_{3}C_{1}, {}_{2}C_{1}, {}_{r}C_{l_3}, {}_{s}C_{k-l_3-2}\right)$ & $1$ & $(N-5)l_3 - r(k-2)$ & $2N-5k$ \\ \hline
$(e'_{R})^c$, $e_{R}$ & $\left({}_{3}C_{3}, {}_{2}C_{0}, {}_{r}C_{l_3}, {}_{s}C_{k-l_3-3}\right)$ & $-6$ & $(N-5)l_3 - r(k-3)$ & $3N-5k$ \\
$(u'_{R})^c$, $u_{R}$ & $\left({}_{3}C_{1}, {}_{2}C_{2}, {}_{r}C_{l_3}, {}_{s}C_{k-l_3-3}\right)$ & $4$ & $(N-5)l_3 - r(k-3)$ & $3N-5k$ \\
$q'_{L}$, $(q_{L})^c$ & $\left({}_{3}C_{2}, {}_{2}C_{1}, {}_{r}C_{l_3}, {}_{s}C_{k-l_3-3}\right)$ & $-1$ & $(N-5)l_3 - r(k-3)$ & $3N-5k$ \\ \hline
$l_{L}$, $(l'_{L})^c$ & $\left({}_{3}C_{3}, {}_{2}C_{1}, {}_{r}C_{l_3}, {}_{s}C_{k-l_3-4}\right)$ & $-3$ & $(N-5)l_3 - r(k-4)$ & $4N-5k$ \\
$(d_{R})^c$, $d'_{R}$ & $\left({}_{3}C_{2}, {}_{2}C_{2}, {}_{r}C_{l_3}, {}_{s}C_{k-l_3-4}\right)$ & $2$ & $(N-5)l_3 - r(k-4)$ & $4N-5k$ \\ \hline
$(\hat{\nu}_{R})^c$, $\nu_{R}$ & $\left({}_{3}C_{3}, {}_{2}C_{2}, {}_{r}C_{l_3}, {}_{s}C_{k-l_3-5}\right)$ & $0$ & $(N-5)l_3 - r(k-5)$ & $5N-5k$ \\ \hline
\end{tabular}
\end{center}
\end{table}
The $U(1)$ charges are those in the following subgroups,
\begin{eqnarray}
\hspace{-5mm} &~& SU(5) \supset  SU(3) \times SU(2) \times U(1)_1 , \\
\hspace{-5mm} &~& SU(N-5) \supset  SU(r) \times SU(N-5-r) \times U(1)_2 ,~ 
	         SU(N-5-1) \times U(1)_2 ,  \\
\hspace{-5mm} &~& SU(N) \supset  SU(5) \times SU(N-5) \times U(1)_3,
\end{eqnarray}
up to normalization.
We assume that $G_{SM} = SU(3) \times SU(2) \times U(1)_1$ up to normalization of the hypercharge.
Particle species are identified with the SM fermions by the gauge quantum numbers.
The $(d_{R})^c$, $l_{L}$, $(u_{R})^c$, $(e_{R})^c$ and $q_{L}$ stand for 
down-type anti-quark singlets, lepton doublets, up-type anti-quark singlets, positron-type lepton singlets and quark doublets. 
The particles with prime are regarded as mirror particles and expected to have no zero modes.
Each fermion has a definite chirality, e.g. $(d_{R})^c$ is left-handed and $d_{R}$ is right-handed.
Here the subscript $L$ ($R$) represents the left-handedness (right-handedness) for Weyl fermions.
The $(d_R)^c$ represents the charge conjugate of $d_R$ and transforms 
as left-handed Weyl fermions under the four-dimensional Lorentz transformation.

A fermion with spin $1/2$ in five dimensions is regarded as a Dirac fermion 
or a pair of Weyl fermions with opposite chiralities in four dimensions.
The left-handed Weyl fermion and the corresponding right-handed one
should have opposite $Z_2$ parity to each other,
from the requirement that the kinetic term is invariant under the $Z_2$ parity transformation.
We define the $Z_2$ parity of the representation~$({}_{p}C_{l_1}, {}_{q}C_{l_2},$ ${}_{r}C_{l_3},$ ${}_{s}C_{l_4})_L$ as follows,
\begin{eqnarray}
\mathcal{P}_0 = (-1)^{l_1+l_2} (-1)^k \eta_k ,~~ 
\mathcal{P}_1 = (-1)^{l_1+l_3} (-1)^k \eta'_k ,
\label{Z2}
\end{eqnarray}
where $\eta_k$ and $\eta'_k$ are the intrinsic $Z_2$ parity.
By definition, $\eta_k$ and $\eta'_k$ take a value $+1$ or $-1$.
We list the $Z_2$ parity assignment for species in Table \ref{t2}.
Note that mirror particles have the $Z_2$ parity $\mathcal{P}_0 = -(-1)^k \eta_{k}$.
Hence {\it all zero modes of mirror particles can be eliminated by a choice of $Z_2$ parity
when we take $(-1)^k \eta_{k} = +1$.}
Hereafter we consider such a case.
\begin{table}
\caption{The $Z_2$ parity assignment for representations of fermions.}
\label{t2}
\begin{center}
\begin{tabular}{c|c|c|c} \hline
species & representation  & $\mathcal{P}_0$ & $\mathcal{P}_1$ \\ \hline\hline
$(\nu_{R})^c$ & $\left({}_{3}C_{0}, {}_{2}C_{0}, {}_{r}C_{l_3}, {}_{s}C_{k-l_3}\right)_L$  
 & $(-1)^k \eta_{k}$ & $(-1)^{l_3} (-1)^k \eta'_{k}$ \\ 
$\hat{\nu}_{R}$ & $\left({}_{3}C_{0}, {}_{2}C_{0}, {}_{r}C_{l_3}, {}_{s}C_{k-l_3}\right)_R$  
 & $-(-1)^k \eta_{k}$ & $-(-1)^{l_3} (-1)^k \eta'_{k}$ \\ \hline
$(d'_{R})^c$ & $\left({}_{3}C_{1}, {}_{2}C_{0}, {}_{r}C_{l_3}, {}_{s}C_{k-l_3-1}\right)_L$  
 & $-(-1)^k \eta_{k}$ & $-(-1)^{l_3} (-1)^k \eta'_{k}$ \\
$l'_{L}$ & $\left({}_{3}C_{0}, {}_{2}C_{1}, {}_{r}C_{l_3}, {}_{s}C_{k-l_3-1}\right)_L$  
 & $-(-1)^k \eta_{k}$  & $(-1)^{l_3} (-1)^k \eta'_{k}$ \\ 
$d_{R}$ & $\left({}_{3}C_{1}, {}_{2}C_{0}, {}_{r}C_{l_3}, {}_{s}C_{k-l_3-1}\right)_R$  
 & $(-1)^k \eta_{k}$ & $(-1)^{l_3} (-1)^k \eta'_{k}$ \\
$(l_{L})^c$ & $\left({}_{3}C_{0}, {}_{2}C_{1}, {}_{r}C_{l_3}, {}_{s}C_{k-l_3-1}\right)_R$  
 & $(-1)^k \eta_{k}$  & $-(-1)^{l_3} (-1)^k \eta'_{k}$ \\ \hline
$(u_{R})^c$ & $\left({}_{3}C_{2}, {}_{2}C_{0}, {}_{r}C_{l_3}, {}_{s}C_{k-l_3-2}\right)_L$ 
 & $(-1)^k \eta_{k}$  & $(-1)^{l_3} (-1)^k \eta'_{k}$ \\
$(e_{R})^c$ & $\left({}_{3}C_{0}, {}_{2}C_{2}, {}_{r}C_{l_3}, {}_{s}C_{k-l_3-2}\right)_L$ 
 & $(-1)^k \eta_{k}$  & $(-1)^{l_3} (-1)^k \eta'_{k}$ \\
$q_{L}$ & $\left({}_{3}C_{1}, {}_{2}C_{1}, {}_{r}C_{l_3}, {}_{s}C_{k-l_3-2}\right)_L$ 
 & $(-1)^k \eta_{k}$  & $-(-1)^{l_3} (-1)^k \eta'_{k}$ \\ 
$u'_{R}$ & $\left({}_{3}C_{2}, {}_{2}C_{0}, {}_{r}C_{l_3}, {}_{s}C_{k-l_3-2}\right)_R$ 
 & $-(-1)^k \eta_{k}$  & $-(-1)^{l_3} (-1)^k \eta'_{k}$ \\
$e'_{R}$ & $\left({}_{3}C_{0}, {}_{2}C_{2}, {}_{r}C_{l_3}, {}_{s}C_{k-l_3-2}\right)_R$ 
 & $-(-1)^k \eta_{k}$  & $-(-1)^{l_3} (-1)^k \eta'_{k}$ \\
$(q'_{L})^c$ & $\left({}_{3}C_{1}, {}_{2}C_{1}, {}_{r}C_{l_3}, {}_{s}C_{k-l_3-2}\right)_R$ 
 & $-(-1)^k \eta_{k}$  & $(-1)^{l_3} (-1)^k \eta'_{k}$ \\ \hline
$(e'_{R})^c$ & $\left({}_{3}C_{3}, {}_{2}C_{0}, {}_{r}C_{l_3}, {}_{s}C_{k-l_3-3}\right)_L$ 
 & $-(-1)^k \eta_{k}$  & $-(-1)^{l_3} (-1)^k \eta'_{k}$ \\
$(u'_{R})^c$ & $\left({}_{3}C_{1}, {}_{2}C_{2}, {}_{r}C_{l_3}, {}_{s}C_{k-l_3-3}\right)_L$ 
 & $-(-1)^k \eta_{k}$  & $-(-1)^{l_3} (-1)^k \eta'_{k}$ \\
$q'_{L}$ & $\left({}_{3}C_{2}, {}_{2}C_{1}, {}_{r}C_{l_3}, {}_{s}C_{k-l_3-3}\right)_L$ 
 & $-(-1)^k \eta_{k}$  & $(-1)^{l_3} (-1)^k \eta'_{k}$ \\ 
$e_{R}$ & $\left({}_{3}C_{3}, {}_{2}C_{0}, {}_{r}C_{l_3}, {}_{s}C_{k-l_3-3}\right)_R$ 
 & $(-1)^k \eta_{k}$  & $(-1)^{l_3} (-1)^k \eta'_{k}$ \\
$u_{R}$ & $\left({}_{3}C_{1}, {}_{2}C_{2}, {}_{r}C_{l_3}, {}_{s}C_{k-l_3-3}\right)_R$ 
 & $(-1)^k \eta_{k}$  & $(-1)^{l_3} (-1)^k \eta'_{k}$ \\
$(q_{L})^c$ & $\left({}_{3}C_{2}, {}_{2}C_{1}, {}_{r}C_{l_3}, {}_{s}C_{k-l_3-3}\right)_R$ 
 & $(-1)^k \eta_{k}$  & $-(-1)^{l_3} (-1)^k \eta'_{k}$ \\ \hline
$l_{L}$ & $\left({}_{3}C_{3}, {}_{2}C_{1}, {}_{r}C_{l_3}, {}_{s}C_{k-l_3-4}\right)_L$ 
 & $(-1)^k \eta_{k}$  & $-(-1)^{l_3} (-1)^k \eta'_{k}$ \\
$(d_{R})^c$ & $\left({}_{3}C_{2}, {}_{2}C_{2}, {}_{r}C_{l_3}, {}_{s}C_{k-l_3-4}\right)_L$ 
 & $(-1)^k \eta_{k}$  & $(-1)^{l_3} (-1)^k \eta'_{k}$ \\ 
$(l'_{L})^c$ & $\left({}_{3}C_{3}, {}_{2}C_{1}, {}_{r}C_{l_3}, {}_{s}C_{k-l_3-4}\right)_R$ 
 & $-(-1)^k \eta_{k}$  & $(-1)^{l_3} (-1)^k \eta'_{k}$ \\
$d'_{R}$ & $\left({}_{3}C_{2}, {}_{2}C_{2}, {}_{r}C_{l_3}, {}_{s}C_{k-l_3-4}\right)_R$ 
 & $-(-1)^k \eta_{k}$  & $-(-1)^{l_3} (-1)^k \eta'_{k}$ \\ \hline
$(\hat{\nu}_{R})^c$ & $\left({}_{3}C_{3}, {}_{2}C_{2}, {}_{r}C_{l_3}, {}_{s}C_{k-l_3-5}\right)_L$ 
 & $-(-1)^k \eta_{k}$  & $-(-1)^{l_3} (-1)^k \eta'_{k}$ \\ 
$\nu_{R}$ & $\left({}_{3}C_{3}, {}_{2}C_{2}, {}_{r}C_{l_3}, {}_{s}C_{k-l_3-5}\right)_R$ 
 & $(-1)^k \eta_{k}$  & $(-1)^{l_3} (-1)^k \eta'_{k}$ \\ \hline
\end{tabular}
\end{center}
\end{table}

We denote the flavor numbers of $(d_{R})^c$, $l_{L}$, $(u_{R})^c$, $(e_{R})^c$, $q_{L}$ 
and (heavy) neutrino singlets as $n_{\bar{d}}$, $n_l$, $n_{\bar{u}}$, $n_{\bar{e}}$, $n_q$ and $n_{\bar{\nu}}$.
Not only left-handed Weyl fermions but also right-handed ones, having even $Z_2$ parities $\mathcal{P}_0 = \mathcal{P}_1 = +1$, 
compose chiral fermions in the SM.
When we take $(-1)^k \eta'_{k} = +1$, the flavor number of each chiral fermions are given by
\begin{align}
n_{\bar{d}} 
	&= \sum_{i = 1, 4} \sum_{l_3 = 0, 2, \dots} {}_{r}C_{l_3} \cdot {}_{N-5-r}C_{k-i-l_3} ,
\label{nd}\\
n_{l}
	&= \sum_{i = 1, 4} \sum_{l_3 = 1, 3, \dots} {}_{r}C_{l_3} \cdot {}_{N-5-r}C_{k-i-l_3} ,
\label{nl}\\
n_{\bar{u}} = n_{\bar{e}} 
	&= \sum_{i = 2, 3} \sum_{l_3 = 0, 2, \dots} {}_{r}C_{l_3} \cdot {}_{N-5-r}C_{k-i-l_3} ,
\label{ne}\\
n_{q} 
	&= \sum_{i = 2, 3} \sum_{l_3 = 1, 3, \dots} {}_{r}C_{l_3} \cdot {}_{N-5-r}C_{k-i-l_3} ,
\label{nq}\\
n_{\bar{\nu}} 
	&= \sum_{i = 0, 5} \sum_{l_3 = 0, 2, \dots} {}_{r}C_{l_3} \cdot {}_{N-5-r}C_{k-i-l_3} ,
\label{nnu}
\end{align}
using the equivalence on charge conjugation.
When we take $(-1)^k \eta'_{k} = -1$, we obtain formulae in which $n_l$ is exchanged by $n_{\bar{d}}$
and $n_q$ by $n_{\bar{u}}$ ($=n_{\bar{e}}$) in Eqs. (\ref{nd}) - (\ref{nq}).
The total number of (heavy) neutrino singlets is given by
$n_{\bar{\nu},k}^{(+-)} = \sum_{i = 0, 5} \sum_{l_3 = 1, 3, \dots} {}_{r}C_{l_3} \cdot {}_{N-5-r}C_{k-i-l_3}$
for $(-1)^k \eta'_{k} = -1$.

For arbitrary $N (\ge 6)$ and $r$, the flavor numbers from $[N, k]$ 
with $((-1)^k \eta_{k},$ $(-1)^k \eta'_{k})$ $= (a, b)$
equal to those from $[N, N-k]$ with $((-1)^{N-k} \eta_{N-k}$, $(-1)^{N-k} \eta'_{N-k})$ $= (a, -b)$ if $r$ is odd
and the flavor numbers from $[N, k]$ 
with $((-1)^k \eta_{k},$ $(-1)^k \eta'_{k})$ $= (a, b)$
equal to those from $[N, N-k]$ with $((-1)^{N-k} \eta_{N-k},$ $(-1)^{N-k} \eta'_{N-k})$ $= (a, b)$ if $r$ is even.
We list the flavor number of each chiral fermion derived from $[N, k]$ ($N = 5, \cdots, 9$ and $k = 1, \cdots, [N/2]$) 
in Table \ref{t3}.
Here $[*]$ stands for Gauss's symbol, i.e., $[N/2] = N/2$ if $N$ is even and $[N/2] = (N-1)/2$ if $N$ is odd.
In the 8-th column, the numbers in the parenthesis are the flavor numbers of neutrino singlets for $(-1)^k \eta'_{k} = -1$.
\begin{table}
\caption{The flavor number of each chiral fermion with $(-1)^k \eta_{k} = (-1)^k \eta'_{k} = +1$.}
\label{t3}
\begin{center}
\begin{tabular}{c|c|c|c|c|c|c|c}
\hline
representation &$(p,q,r,s)$&$n_{\bar{d}}$&$n_{l}$&$n_{\bar{u}}$&$n_{\bar{e}}$&$n_{q}$&$n_{\bar{\nu}}$ 
($n_{\bar{\nu}}$ with $(-1)^k \eta'_{k} = -1$) \\
\hline
$[N,1]$&$(3,2,r,s)$&1&0&0&0&0&$s~(r)$\\
\hline
$[N,2]$&$(3,2,r,s)$&$s$&$r$&1&1&0&${}_rC_2+{}_sC_2$~$(rs)$\\
\hline
$[6,3]$&(3,2,1,0)&0&0&1&1&1&0~(0)\\
\cline{2-8}
&(3,2,0,1)&0&0&2&2&0&0~(0)\\
\hline
&(3,2,2,0)&1&0&1&1&2&0~(0)\\
\cline{2-8}
$[7,3]$&(3,2,1,1)&0&1&2&2&1&0~(0)\\
\cline{2-8}
&(3,2,0,2)&1&0&3&3&0&0~(0)\\
\hline
&(3,2,3,0)&3&0&1&1&3&0~(1)\\
\cline{2-8}
$[8,3]$&(3,2,2,1)&1&2&2&2&2&1~(0)\\
\cline{2-8}
&(3,2,1,2)&1&2&3&3&1&0~(1)\\
\cline{2-8}
&(3,2,0,3)&3&0&4&4&0&1~(0)\\
\hline
&(3,2,3,0)&1&1&3&3&3&0~(0)\\
\cline{2-8}
$[8,4]$&(3,2,2,1)&2&0&2&2&4&0~(0)\\
\cline{2-8}
&(3,2,1,2)&1&1&3&3&3&0~(0)\\
\cline{2-8}
&(3,2,0,3)&2&0&6&6&0&0~(0)\\
\hline
&(3,2,4,0)&6&0&1&1&4&0~(4)\\
\cline{2-8}
&(3,2,3,1)&3&3&2&2&3&3~(1)\\
\cline{2-8}
$[9,3]$&(3,2,2,2)&2&4&3&3&2&2~(2)\\
\cline{2-8}
&(3,2,1,3)&3&3&4&4&1&1~(3)\\
\cline{2-8}
&(3,2,0,4)&6&0&5&5&0&4~(0)\\
\hline
&(3,2,4,0)&1&4&6&6&4&1~(0)\\
\cline{2-8}
&(3,2,3,1)&4&1&4&4&6&0~(1)\\
\cline{2-8}
$[9,4]$&(3,2,2,2)&3&2&4&4&6&1~(0)\\
\cline{2-8}
&(3,2,1,3)&2&3&6&6&4&0~(1)\\
\cline{2-8}
&(3,2,0,4)&5&0&10&10&0&1~(0)\\
\hline
\end{tabular}
\end{center}
\end{table}

Our four-dimensional world is assumed to be a boundary at one of the fixed points, on the basis of the `brane world scenario'.
There exist two kinds of four-dimensional field in our low-energy theory.
One is the brane field which lives only at the boundary and the other is the zero mode stemming from the bulk field.
The Kaluza-Klein (KK) modes do not appear in our low-energy world 
because they have heavy masses of $O(1/R)$, the magnitude same as the unification scale $M_U$.
There are many possibilities to derive three families from zero modes of (a few of) bulk field
and suitable brane fields in the view point of chiral anomaly cancellation.
Chiral anomalies may arise at the boundaries with the advent of chiral fermions.
Those anomalies must be cancelled in the four-dimensional effective theory by the contribution of brane chiral fermions
and/or counter terms such as the Chern-Simons term~\cite{C&H,AC&G,KK&L}.

\section{Sfermion mass relations}

We consider the SUSY version of $SU(N)$ models.
In SUSY models, the hypermultiplet in the five-dimensional bulk is equivalent to a pair of chiral multiplets 
with opposite gauge quantum numbers in four dimensions.
The chiral multiplet with the representation $[N,N-k]$, which is a conjugate of $[N,k]$, 
contains a left-handed Weyl fermion with $[N,N-k]_L$.
This Weyl fermion is regarded as a right-handed one with $[N,k]_R$ by using the charge conjugation.
Hence our analysis in the previous section works on SUSY models.

We take the following assumptions in our analysis.\\
1. Three families in the MSSM come from zero modes of the bulk field 
with the representation $[N, k]$ and some brane fields.\\
2. We do not specifiy the mechanism that the $N=1$ SUSY is broken down in four dimensions.\footnote{
Scherk-Schwarz mechanism, in which SUSY is broken by the difference of BCs 
between bosons and fermions, is a typical one.\cite{S&S}
This mechanism on $S^1/Z_2$ leads to a restricted type of soft SUSY breaking parameters such as
$M_i = \beta/R$ for bulk gauginos and $m_{\tilde{f}}^2 = (\beta/R)^2$ for bulk scalar particles
where $\beta$ is a real parameter and $R$ is a radius of $S^1$.}
Soft SUSY breaking terms respect the gauge invariance.\\
3. Extra gauge symmetries are broken by Higgs mechanism at the same time as the orbifold breaking at the scale $M_U = O(1/R)$.
Then the $D$-term contributions to scalar masses can appear as a dominant source of scalar mass splitting.


\subsection{Sfermion mass formula}

We consider the case with the intrinsic $Z_2$ parity assignment $(-1)^k \eta_{k} = (-1)^k \eta'_{k} = +1$.
In the case with $(-1)^k \eta'_{k} = -1$, similar relations are derived by a suitable exchange of sfermion species.
We list sfermion species as zero modes of five-dimensional fields with even $Z_2$ parities $\mathcal{P}_0 = \mathcal{P}_1 = +1$
in Table \ref{t4}.
\begin{table}
\caption{The assignment of sfermions and those $U(1)$ charges.}
\label{t4}
\begin{center}
\begin{tabular}{c|c|c|c|c} \hline
species & $(l_1, l_2, l_3)$  & $l_1 + l_2$ & $U(1)_2$ & $U(1)_3$ \\ \hline\hline
$\tilde{d}^*_{R}$ & $(1, 0, {\mbox{even}})$ & $1$ & $-(N-5)l_3 + r(k-1)$ & $-N+5k$ \\
$\tilde{l}_{L}$ & $(0, 1, {\mbox{odd}})$  & $1$ & $-(N-5)l_3 + r(k-1)$ & $-N+5k$ \\ \hline
$\tilde{u}^*_{R}$ & $(2, 0, {\mbox{even}})$ & $2$ & $(N-5)l_3 - r(k-2)$ & $2N-5k$ \\
$\tilde{e}^*_{R}$ & $(0, 2, {\mbox{even}})$ & $2$ & $(N-5)l_3 - r(k-2)$ & $2N-5k$ \\
$\tilde{q}_{L}$ & $(1, 1, {\mbox{odd}})$ & $2$ & $(N-5)l_3 - r(k-2)$ & $2N-5k$ \\ \hline
$\tilde{e}^*_{R}$ & $(3, 0, {\mbox{even}})$ & $3$ & $-(N-5)l_3 + r(k-3)$ & $-3N+5k$ \\
$\tilde{u}^*_{R}$ & $(1, 2, {\mbox{even}})$ & $3$ & $-(N-5)l_3 + r(k-3)$ & $-3N+5k$ \\
$\tilde{q}_{L}$ & $(2, 1, {\mbox{odd}})$ & $3$ & $-(N-5)l_3 + r(k-3)$ & $-3N+5k$ \\ \hline
$\tilde{l}_{L}$ & $(3, 1, {\mbox{odd}})$ & $4$ & $(N-5)l_3 - r(k-4)$ & $4N-5k$ \\
$\tilde{d}^*_{R}$ & $(2, 2, {\mbox{even}})$ & $4$ & $(N-5)l_3 - r(k-4)$ & $4N-5k$ \\ \hline
\end{tabular}
\end{center}
\end{table}
In Table \ref{t4}, $\tilde{f}$ means the scalar partner of fermion $f$,
and the charge-conjugation is performed for the fields with $l_1 + l_2 =$ odd.
The asterisk stands for the complex conjugate.
Note that the sign of $U(1)$ charges is changed by the charge-conjugation.
As sfermion species are labeled by the numbers ($l_1, l_2, l_3$),
we use this label in place of $\tilde{f}$.

Sfermion mass squareds at $M_U$ are written by
\begin{eqnarray}
&~& {m_{(l_1,l_2,l_3)}^{(\alpha, \beta)}}^2(M_U) = m_{[N,k]}^2 + (-1)^{l_1+l_2}\sum_{A=1}^{r-1} Q_{\alpha}^A D_{(r)}^A
 + (-1)^{l_1+l_2}\sum_{B=1}^{r-1} Q_{\beta}^B D_{(s)}^B
\nonumber\\
&~&~~~~~~~~~~~~~~~~~~~~ + (-1)^{l_1+l_2}\left[(N-5)l_3-r(k-(l_1+l_2))\right]D_2 
\nonumber\\
&~&~~~~~~~~~~~~~~~~~~~~ + (-1)^{l_1+l_2}\left[(l_1+l_2)N-5k\right]D_3 ,
\label{sf-mass}
\end{eqnarray}
where $m_{[N,k]}^2$ is a common soft SUSY breaking mass parameter which respects $SU(N)$ gauge symmetry
and other terms in the right-hand side represent $D$-term contributions.
The $D$-term contributions, in general, originate from $D$-terms related to broken gauge symmetries 
when soft SUSY breaking parameters possess a non-universal structure and
the rank of gauge group lowers after the breakdown of gauge symmetry.\cite{Dterm,KM&Y}
In most cases, the magnitude of $D$-term condensation is, at most, of order TeV scale squared
and hence $D$-term contributions can induce sizable effects on sfermion spectrum.
The $\alpha$ and $\beta$ represent indices which indicate members of multiplet of $SU(r)$ and $SU(s)$, 
and run from 1 to ${}_rC_{l_3}$ and from 1 to ${}_sC_{l_4}$, respectively.
The $Q_{\alpha}^A$ are broken diagonal charges of $[r,l_3]$, which form the Cartan sub-algebra of $SU(r)$,
and given by
\begin{eqnarray}
Q_{\alpha}^A = Q_{\alpha}^A([r, l_3]) = \sum_{a=a_1}^{a_{l_3}} Q_{a}^A ,
\label{QA}
\end{eqnarray}
where $Q_{a}^A$ are the diagonal charges (up to normalization) for fields with the fundamental representation $[r,1]$ defined by
\begin{eqnarray}
Q_{a}^A \equiv (1-a) \delta_{a-1}^A + \sum_{i=0}^{r-1-a} \delta_{a+i}^A .
\label{QaA}
\end{eqnarray}
The numbering for $\alpha$ is defined by
\begin{align}
\hspace{-5mm} (a_1, \cdots, a_{l_3}) &= (1, \cdots, l_3) ~~~~~~~~~~~~~ &\mbox{for}~ \alpha=1 
\nonumber \\
\hspace{-5mm}   &= (1, \cdots, l_3-1, l_3+1)  &\mbox{for}~ \alpha=2 
\nonumber \\
\hspace{-5mm}   ~~~~~~ & \cdots 
\nonumber \\
\hspace{-5mm}  &= (1, \cdots, l_3-1, r)  ~~~~~~~~~~~&\mbox{for}~ \alpha=l_3-r+1 
\nonumber \\
\hspace{-5mm}  &= (1, \cdots, l_3-2, l_3, l_3+1) &\mbox{for}~ \alpha=l_3-r+2 
\nonumber \\
\hspace{-5mm}  ~~~~~~ & \cdots 
\nonumber \\
\hspace{-5mm}  &= (r+1-l_3, \cdots, r)  &\mbox{for}~ \alpha={}_rC_{l_3} .
\label{alpha}
\end{align}
By using formulae of diagonal charges (\ref{QA}) and (\ref{QaA}) and the definition of numbering (\ref{alpha}), 
the broken diagonal charges of $[r,r-l_3]$ (the complex conjugate representation of $[r,l_3]$) are given by
\begin{eqnarray}
Q_{\alpha}^A([r,r-l_3]) = -Q_{{}_rC_{l_3}+1-\alpha}^A([r, l_3]) .
\label{QA-cc}
\end{eqnarray}
The same holds on the charges $Q_{\beta}^B$.
The $D_{(r)}^A$, $D_{(s)}^B$, $D_2$ and $D_3$ are parameters including $D$-term condensations
and those magnitudes are model-dependent.

\subsection{Sfermion mass relations}

Let us derive relations among sfermion masses at $M_U$, by eliminating unknown parameters 
($m_{[N,k]}^2$, $D_{(r)}^A$, $D_{(s)}^B$, $D_2$, $D_3$) in the mass formula (\ref{sf-mass}). 

First of all, we find the following relations from the mass formula (\ref{sf-mass}) and Table \ref{t4},
\begin{eqnarray}
{m_{(2,0,l_3)}^{(\alpha, \beta)}}^2 = {m_{(0,2,l_3)}^{(\alpha, \beta)}}^2 ,~
{m_{(3,0,l_3)}^{(\alpha, \beta)}}^2 = {m_{(1,2,l_3)}^{(\alpha, \beta)}}^2 .
\label{rel-1}
\end{eqnarray}
Here and hereafter we abbreviate ${m_{(l_1,l_2,l_3)}^{(\alpha, \beta)}}^2(M_U)$ as ${m_{(l_1,l_2,l_3)}^{(\alpha, \beta)}}^2$.
This type of relation generally appears if up-type anti-squark singlet exists, 
and the number of relations is $n_{\bar{u}} (= n_{\bar{e}})$.
Hereafter we consider only up-type anti-squark singlets (in place of positron-type slepton singlets).

Before we derive other relations, we estimate total number of independent relations. 
Number of each sfermion derived from bulk field $[N,k]$ equals to 
that of each fermion given in Eqs. (\ref{nd}) - (\ref{nnu}). 
Total number of sfermions excluding slepton singlets is given by 
\begin{eqnarray}
N_{\tiny{\mbox{tot}}} = \sum_{i=1}^{4}\sum_{l_3=0,1,\ldots}{}_rC_{l_3}\cdot {}_{N-5-r}C_{k-i-l_3}  = \sum_{i=1}^4{}_{N-5}C_{k-i} .
\end{eqnarray}
The number of unknown parameters is $N-4$ because the number of $D$-term condensations equals to 
the difference of rank between $SU(N)$ and $G_{SM}$. 
Hence the number of independent relations excluding (\ref{rel-1}) is $N_{\tiny{\mbox{tot}}} - N + 4$.
We find that no relation is derived from $[N,1]$ 
and one relation $\displaystyle{{m_{(2,0,0)}^{(\alpha, \beta)}}^2 = {m_{(0,2,0)}^{(\alpha, \beta)}}^2}$ (the type (\ref{rel-1})) from $[N,2]$.

By taking the summation over all members in each multiplet 
of $SU(r)$ and $SU(s)$, the following formula is derived,
\begin{eqnarray}
&~& \sum_{\alpha=1}^{{}_rC_{l_3}} \sum_{\beta=1}^{{}_sC_{l_4}} {m_{(l_1,l_2,l_3)}^{(\alpha, \beta)}}^2
\nonumber \\
&~& ~~~~ \left. = {}_rC_{l_3} \cdot {}_sC_{l_4} \left(m_{[N,k]}^2\right.
+ (-1)^{l_1+l_2}\left[(N-5)l_3-r(k-(l_1+l_2))\right]D_2\right.
\nonumber \\
&~& ~~~~~~ \left. + (-1)^{l_1+l_2}\left[(l_1+l_2)N-5k\right]D_3\right) ,
\label{sf-mass-sum}
\end{eqnarray}
Note that both $D_{(r)}^A$ and $D_{(s)}^B$ disappear because of the traceless property of diagonal generators.
If the number of multiplets ($N_{\tiny{\mbox{mul}}}$) is beyond three, $N_{\tiny{\mbox{mul}}} - 3$ kinds of relations are 
derived by eliminating unknown parameters ($m_{[N,k]}^2$, $D_2$, $D_3$).

The remaining relations are derived by a summation among multiplets with suitable coefficients (not a universal one), 
and are formally written by
\begin{eqnarray}
\hspace{-5mm} \sum_{\alpha} c_{\alpha} {m_{(l_1,l_2,l_3)}^{(\alpha, \beta)}}^2 
= \sum_{\alpha'} c'_{\alpha'} {m_{(l'_1,l'_2,l'_3)}^{(\alpha', \beta')}}^2  ,~
\sum_{\beta} d_{\beta} {m_{(l_1,l_2,l_3)}^{(\alpha, \beta)}}^2 
= \sum_{\beta'} d'_{\beta'} {m_{(l'_1,l'_2,l'_3)}^{(\alpha', \beta')}}^2  ,
\label{sf-mass-multiplet}
\end{eqnarray}
where $c_{\alpha}$, $c'_{\alpha}$, $d_{\beta}$ and $d'_{\beta}$ are coefficients which satisfies the following 
relations,
\begin{eqnarray}
&~& \sum_{\alpha} c_{\alpha} = \sum_{\alpha'} c'_{\alpha'} ,~
\sum_{\alpha} c_{\alpha} Q_{\alpha}^A = \sum_{\alpha'} c'_{\alpha'} Q_{\alpha'}^A ,
\label{c}\\
&~& \sum_{\beta} d_{\beta} = \sum_{\beta'} d'_{\beta'} ,~
\sum_{\beta} d_{\beta} Q_{\beta}^A = \sum_{\beta'} d'_{\beta'} Q_{\beta'}^A .
\label{d}
\end{eqnarray}

Sfermion mass relations (excluding the type (\ref{rel-1})) derived from $[6,3]$ - $[9,4]$ are listed in Table \ref{t5}. 
\begin{table}
\caption{The sfermion mass relations derived from $[6,3]$ - $[9,4]$.}
\label{t5}
\begin{center}
\begin{tabular}{c|c|l}
\hline
rep. & $(p,q,r,s)$ &~~~~~~~~~~~~~~~~~~~~~~~~sfermion mass relations\\
\hline
$[6,3]$&(3,2,1,0)& ${m_{(1,1,1)}^{(1,1)}}^2={m_{(1,2,0)}^{(1,1)}}^2$\\
\cline{2-3}
&(3,2,0,1)&${m_{(2,0,0)}^{(1,1)}}^2={m_{(1,2,0)}^{(1,1)}}^2$\\
\hline
$[7,3]$&(3,2,2,0)&$\displaystyle{5{m_{(1,0,2)}^{(1,1)}}^2+9{m_{(1,2,0)}^{(1,1)}}^2=7\sum_{\alpha=1}^2{m_{(1,1,1)}^{(\alpha,1)}}^2}$\\
\cline{2-3}
&(3,2,1,1)&$\displaystyle{5{m_{(0,1,1)}^{(1,1)}}^2+9{m_{(1,2,0)}^{(1,1)}}^2=7\left({m_{(1,1,1)}^{(1,1)}}^2+{m_{(2,0,0)}^{(1,1)}}^2\right)}$\\
\cline{2-3}
&(3,2,0,2)&$\displaystyle{5{m_{(1,0,0)}^{(1,1)}}^2+9{m_{(1,2,0)}^{(1,1)}}^2=7\sum_{\beta=1}^2{m_{(2,0,0)}^{(1,\beta)}}^2}$\\
\hline
&(3,2,3,0)&$\displaystyle{5\sum_{\alpha=1}^3{m_{(1,0,2)}^{(\alpha,1)}}^2 +9{m_{(1,2,0)}^{(1,1)}}^2 =8\sum_{\alpha=1}^3{m_{(1,1,1)}^{(\alpha,1)}}^2}$,\\
&&${m_{(1,0,2)}^{(3,1)}}^2-{m_{(1,1,1)}^{(1,1)}}^2={m_{(1,0,2)}^{(2,1)}}^2-{m_{(1,1,1)}^{(2,1)}}^2={m_{(1,0,2)}^{(1,1)}}^2-{m_{(1,1,1)}^{(3,1)}}^2$\\
\cline{2-3}
&(3,2,2,1)&$\displaystyle{\sum_{\alpha =1}^2{m_{(1,1,1)}^{(\alpha,1)}}^2+2{m_{(1,0,2)}^{(1,1)}}^2
 =\sum_{\alpha=1}^2{m_{(0,1,1)}^{(\alpha,1)}}^2+2{m_{(2,0,0)}^{(1,1)}}^2}$,\\
&& $\displaystyle{6{m_{(2,0,0)}^{(1,1)}}^2 + \sum_{\alpha =1}^2{m_{(1,1,1)}^{(\alpha,1)}}^2 
 = 5{m_{(1,0,2)}^{(1,1)}}^2+3{m_{(1,2,0)}^{(1,1)}}^2}$,\\
$[8,3]$&&${m_{(0,1,1)}^{(1,1)}}^2-{m_{(0,1,1)}^{(2,1)}}^2={m_{(1,1,1)}^{(2,1)}}^2-{m_{(1,1,1)}^{(1,1)}}^2$\\
\cline{2-3}
&(3,2,1,2)&$\displaystyle{\sum_{\beta=1}^2{m_{(2,0,0)}^{(1,\beta)}}^2+2{m_{(1,0,0)}^{(1,1)}}^2
=\sum_{\beta=1}^2{m_{(0,1,1)}^{(1,\beta)}}^2+2{m_{(1,1,1)}^{(1,1)}}^2}$,\\
&& $\displaystyle{6{m_{(1,1,1)}^{(1,1)}}^2 + \sum_{\beta =1}^2{m_{(2,0,0)}^{(1,\beta)}}^2 
 = 5{m_{(1,0,0)}^{(1,1)}}^2+3{m_{(1,2,0)}^{(1,1)}}^2}$,\\
&&${m_{(0,1,1)}^{(1,1)}}-{m_{(0,1,1)}^{(1,2)}}={m_{(2,0,0)}^{(1,2)}}-{m_{(2,0,0)}^{(1,1)}}$\\
\cline{2-3}
&(3,2,0,3)&$\displaystyle{5\sum_{\beta=1}^3{m_{(1,0,0)}^{(1,\beta)}}^2+9{m_{(1,2,0)}^{(1,1)}}^2=8\sum_{\beta=1}^3{m_{(2,0,0)}^{(1,\beta)}}^2}$,\\
&&${m_{(1,0,0)}^{(1,3)}}^2-{m_{(2,0,0)}^{(1,1)}}^2={m_{(1,0,0)}^{(1,2)}}^2-{m_{(2,0,0)}^{(1,2)}}^2
={m_{(1,0,0)}^{(1,1)}}^2-{m_{(2,0,0)}^{(1,3)}}^2$\\
\hline
&(3,2,3,0)& ${m_{(0,1,3)}^{(1,1)}}^2={m_{(2,2,0)}^{(1,1)}}^2$,\\
&&${m_{(2,0,2)}^{(1,1)}}^2= {m_{(2,1,1)}^{(3,1)}}^2$,~ ${m_{(2,0,2)}^{(2,1)}}^2 = {m_{(2,1,1)}^{(2,1)}}^2$,~ 
${m_{(2,0,2)}^{(3,1)}}^2 = {m_{(2,1,1)}^{(1,1)}}^2$\\
\cline{2-3}
&(3,2,2,1)& ${m_{(1,0,2)}^{(1,1)}}^2={m_{(2,2,0)}^{(1,1)}}^2$,~ ${m_{(2,0,2)}^{(1,1)}}^2={m_{(1,2,0)}^{(1,1)}}^2$,\\
$[8,4]$&& ${m_{(1,1,1)}^{(1,1)}}^2={m_{(2,1,1)}^{(2,1)}}^2$,~ ${m_{(1,1,1)}^{(2,1)}}^2={m_{(2,1,1)}^{(1,1)}}^2$\\
\cline{2-3}
&(3,2,1,2)& ${m_{(0,1,1)}^{(1,1)}}^2={m_{(2,2,0)}^{(1,1)}}^2$,~ ${m_{(2,0,0)}^{(1,1)}}^2={m_{(2,1,1)}^{(1,1)}}^2$,\\
&& ${m_{(1,1,1)}^{(1,1)}}^2={m_{(1,2,0)}^{(1,2)}}^2$,
~ ${m_{(1,1,1)}^{(1,2)}}^2={m_{(1,2,0)}^{(1,1)}}^2$\\
\cline{2-3}
&(3,2,0,3)&${m_{(1,0,0)}^{(1,1)}}^2={m_{(2,2,0)}^{(1,1)}}^2$,\\
&&${m_{(2,0,0)}^{(1,1)}}^2= {m_{(1,2,0)}^{(1,3)}}^2$,~ ${m_{(2,0,0)}^{(1,2)}}^2 = {m_{(1,2,0)}^{(1,2)}}^2$,~ 
${m_{(2,0,0)}^{(1,3)}}^2 = {m_{(1,2,0)}^{(1,1)}}^2$\\
\hline
\end{tabular}
\end{center}
\end{table}
\begin{table}
\begin{center}
\begin{tabular}{c|c|l}
\hline
rep.&$(p,q,r,s)$&~~~~~~~~~~~~~~~~~~~~~~~~~~~~~~~~~~sfermion mass relations\\
\hline
&(3,2,4,0)&$\displaystyle{9\sum_{\alpha =1}^4{m_{(1,1,1)}^{(\alpha,1)}}^2=5\sum_{\alpha=1}^6{m_{(1,0,2)}^{(\alpha,1)}}^2+6{m_{(1,2,0)}^{(1,1)}}^2}$,\\
&&${m_{(1,1,1)}^{(1,1)}}^2-{m_{(1,1,1)}^{(2,1)}}^2={m_{(1,0,2)}^{(4,1)}}^2-{m_{(1,0,2)}^{(2,1)}}^2
={m_{(1,0,2)}^{(5,1)}}^2-{m_{(1,0,2)}^{(3,1)}}^2$,\\
&&${m_{(1,1,1)}^{(1,1)}}^2-{m_{(1,1,1)}^{(3,1)}}^2={m_{(1,0,2)}^{(4,1)}}^2-{m_{(1,0,2)}^{(1,1)}}^2={m_{(1,0,2)}^{(6,1)}}^2-{m_{(1,0,2)}^{(3,1)}}^2$,\\
&&${m_{(1,1,1)}^{(1,1)}}^2-{m_{(1,1,1)}^{(4,1)}}^2={m_{(1,0,2)}^{(5,1)}}^2-{m_{(1,0,2)}^{(1,1)}}^2$\\
\cline{2-3}
&(3,2,3,1)&$\displaystyle{\sum_{\alpha=1}^3{m_{(1,0,2)}^{(\alpha,1)}}^2+\sum_{\alpha=1}^3{m_{(1,1,1)}^{(\alpha,1)}}^2
=\sum_{\alpha=1}^3{m_{(0,1,1)}^{(\alpha,1)}}^2+3{m_{(2,0,0)}^{(1,1)}}^2}$,\\
&&$\displaystyle{\sum_{\alpha =1}^3{m_{(1,0,2)}^{(\alpha,1)}}^2 + 4\sum_{\alpha=1}^3{m_{(0,1,1)}^{(\alpha,1)}}^2
 + 3{m_{(1,2,0)}^{(1,1)}}^2 = 6\sum_{\alpha=1}^3 {m_{(1,1,1)}^{(\alpha,1)}}^2}$,\\
&&${m_{(1,1,1)}^{(1,1)}}^2-{m_{(1,1,1)}^{(2,1)}}^2={m_{(1,0,2)}^{(3,1)}}^2-{m_{(1,0,2)}^{(2,1)}}^2={m_{(0,1,1)}^{(2,1)}}^2-{m_{(0,1,1)}^{(1,1)}}^2$,\\
&&${m_{(1,1,1)}^{(1,1)}}^2-{m_{(1,1,1)}^{(3,1)}}^2={m_{(1,0,2)}^{(3,1)}}^2-{m_{(1,0,2)}^{(1,1)}}^2={m_{(0,1,1)}^{(3,1)}}^2-{m_{(0,1,1)}^{(1,1)}}^2$\\
\cline{2-3}
&(3,2,2,2)&$\displaystyle{{m_{(1,0,2)}^{(1,1)}}^2 + \sum_{\alpha=1}^{2}{m_{(1,1,1)}^{(\alpha,1)}}^2
= {m_{(1,0,0)}^{(1,1)}}^2 + \sum_{\beta=1}^2{m_{(2,0,0)}^{(1,\beta)}}^2}$,\\
&& $\displaystyle{3\left(\sum_{\alpha=1}^{2}{m_{(1,1,1)}^{(\alpha,1)}}^2 + \sum_{\beta=1}^{2}{m_{(2,0,0)}^{(1,\beta)}}^2\right)
= 2{m_{(1,2,0)}^{(1,1)}}^2 + 5\left({m_{(1,0,0)}^{(1,1)}}^2+ {m_{(1,0,2)}^{(1,1)}}^2\right)}$,\\
$[9,3]$&&$\displaystyle{2{m_{(1,0,0)}^{(1,1)}}^2+2{m_{(1,0,2)}^{(1,1)}}^2=\sum_{\alpha=1}^2\sum_{\beta=1}^2{m_{(0,1,1)}^{(\alpha,\beta)}}^2}$,\\
&&${m_{(1,1,1)}^{(1,1)}}^2-{m_{(1,1,1)}^{(2,1)}}^2={m_{(0,1,1)}^{(2,1)}}^2-{m_{(0,1,1)}^{(1,1)}}^2
 = {m_{(0,1,1)}^{(2,2)}}^2-{m_{(0,1,1)}^{(1,2)}}^2$,\\
&&${m_{(2,0,0)}^{(1,1)}}^2-{m_{(2,0,0)}^{(1,2)}}^2={m_{(0,1,1)}^{(1,2)}}^2-{m_{(0,1,1)}^{(1,1)}}^2$\\
\cline{2-3}
&(3,2,1,3)&$\displaystyle{\sum_{\beta=1}^3{m_{(1,0,0)}^{(1,\beta)}}^2+\sum_{\beta=1}^3{m_{(2,0,0)}^{(1,\beta)}}^2
=\sum_{\beta=1}^3{m_{(0,1,1)}^{(1,\beta)}}^2+3{m_{(1,1,1)}^{(1,1)}}^2}$,\\
&&$\displaystyle{\sum_{\beta =1}^3{m_{(1,0,0)}^{(1,\beta)}}^2 + 4\sum_{\beta=1}^3{m_{(0,1,1)}^{(1,\beta)}}^2
 + 3{m_{(1,2,0)}^{(1,1)}}^2 = 6\sum_{\beta=1}^3 {m_{(2,0,0)}^{(1,\beta)}}^2}$,\\
&&${m_{(2,0,0)}^{(1,1)}}^2-{m_{(2,0,0)}^{(1,2)}}^2={m_{(1,0,0)}^{(1,3)}}^2-{m_{(1,0,0)}^{(1,2)}}^2={m_{(0,1,1)}^{(1,2)}}^2-{m_{(0,1,1)}^{(1,1)}}^2$,\\
&&${m_{(2,0,0)}^{(1,1)}}^2-{m_{(2,0,0)}^{(1,3)}}^2={m_{(1,0,0)}^{(1,3)}}^2-{m_{(1,0,0)}^{(1,1)}}^2={m_{(0,1,1)}^{(1,3)}}^2-{m_{(0,1,1)}^{(1,1)}}^2$\\
\cline{2-3}
&(3,2,0,4)&$\displaystyle{9\sum_{\beta =1}^4{m_{(2,0,0)}^{(1,\beta)}}^2=5\sum_{\beta=1}^6{m_{(1,0,0)}^{(1,\beta)}}^2+6{m_{(1,2,0)}^{(1,1)}}^2}$,\\
&&${m_{(2,0,0)}^{(1,1)}}^2-{m_{(2,0,0)}^{(1,2)}}^2={m_{(1,0,0)}^{(1,4)}}^2-{m_{(1,0,0)}^{(1,2)}}^2={m_{(1,0,0)}^{(1,5)}}^2-{m_{(1,0,0)}^{(1,3)}}^2$,\\
&&${m_{(2,0,0)}^{(1,1)}}^2-{m_{(2,0,0)}^{(1,3)}}^2={m_{(1,0,0)}^{(1,4)}}^2-{m_{(1,0,0)}^{(1,1)}}^2={m_{(1,0,0)}^{(1,6)}}^2-{m_{(1,0,0)}^{(1,3)}}^2$,\\
&&${m_{(2,0,0)}^{(1,1)}}^2-{m_{(2,0,0)}^{(1,4)}}^2={m_{(1,0,0)}^{(1,5)}}^2-{m_{(1,0,0)}^{(1,1)}}^2$\\
\hline
\end{tabular}
\end{center}
\end{table}
\begin{table}
\begin{center}
\begin{tabular}{c|c|l}
\hline
rep.&$(p,q,r,s)$&~~~~~~~~~~~~~~~~~~~~~~~~~~sfermion mass relations\\
\hline
&(3,2,4,0)&$\displaystyle{12\sum_{\alpha=1}^6{m_{(2,0,2)}^{(\alpha,1)}}^2
=5\sum_{\alpha=1}^4{m_{(0,1,3)}^{(\alpha,1)}}^2+13\sum_{\alpha=1}^4{m_{(2,1,1)}^{(\alpha,1)}}^2}$,\\
&&$\displaystyle{23\sum_{\alpha=1}^6{m_{(2,0,2)}^{(\alpha,1)}}^2=27\sum_{\alpha=1}^4{m_{(2,1,1)}^{(\alpha,1)}}^2+30{m_{(2,2,0)}^{(1,1)}}^2}$,\\
&&${m_{(0,1,3)}^{(2,1)}}^2-{m_{(0,1,3)}^{(1,1)}}^2={m_{(2,1,1)}^{(4,1)}}^2-{m_{(2,1,1)}^{(3,1)}}^2$\\
&&$~~~={m_{(2,0,2)}^{(2,1)}}^2-{m_{(2,0,2)}^{(4,1)}}^2
={m_{(2,0,2)}^{(3,1)}}^2-{m_{(2,0,2)}^{(5,1)}}^2$,\\
&&${m_{(0,1,3)}^{(3,1)}}^2-{m_{(0,1,3)}^{(1,1)}}^2={m_{(2,1,1)}^{(4,1)}}^2-{m_{(2,1,1)}^{(2,1)}}^2$\\
&&$~~~={m_{(2,0,2)}^{(1,1)}}^2-{m_{(2,0,2)}^{(4,1)}}^2
={m_{(2,0,2)}^{(3,1)}}^2-{m_{(2,0,2)}^{(6,1)}}^2$,\\
&&${m_{(0,1,3)}^{(4,1)}}^2-{m_{(0,1,3)}^{(1,1)}}^2={m_{(2,1,1)}^{(4,1)}}^2-{m_{(2,1,1)}^{(1,1)}}^2$\\
&&$~~~={m_{(2,0,2)}^{(1,1)}}^2-{m_{(2,0,2)}^{(5,1)}}^2$\\
\cline{2-3}
&(3,2,3,1)&$\displaystyle{\sum_{\alpha=1}^3{m_{(1,0,2)}^{(\alpha,1)}}^2-3{m_{(0,1,3)}^{(1,1)}}^2
=\sum_{\alpha=1}^3{m_{(2,0,2)}^{(\alpha,1)}}^2-\sum_{\alpha=1}^3{m_{(1,1,1)}^{(\alpha,1)}}^2}$\\
&&$~~~\displaystyle{=3{m_{(1,2,0)}^{(1,1)}}^2-\sum_{\alpha=1}^3{m_{(2,1,1)}^{(\alpha,1)}}^2}$,\\
&&$\displaystyle{27\left({m_{(1,2,0)}^{(1,1)}}^2+{m_{(0,1,3)}^{(1,1)}}^2\right)
 = 12{m_{(2,2,0)}^{(1,1)}}^2 + 7\left(\sum_{\alpha=1}^3{m_{(2,0,2)}^{(\alpha,1)}}^2+\sum_{\alpha=1}^3{m_{(1,1,1)}^{(\alpha,1)}}^2\right)}$,\\
$[9,4]$&&$\displaystyle{8 \sum_{\alpha=1}^3{m_{(1,0,2)}^{(\alpha,1)}}^2+\sum_{\alpha=1}^3{m_{(2,1,1)}^{(\alpha,1)}}^2}$,\\
&&$~~~\displaystyle{= {m_{(0,1,3)}^{(1,1)}}^2+8{m_{(1,2,0)}^{(1,1)}}^2+18{m_{(2,2,0)}^{(1,1)}}^2}$\\
&&${m_{(1,0,2)}^{(1,1)}}^2-{m_{(1,0,2)}^{(2,1)}}^2={m_{(1,1,1)}^{(3,1)}}^2-{m_{(1,1,1)}^{(2,1)}}^2$\\
&&$~~~={m_{(2,1,1)}^{(2,1)}}^2-{m_{(2,1,1)}^{(3,1)}}^2= {m_{(2,0,2)}^{(2,1)}}^2-{m_{(2,0,2)}^{(1,1)}}^2$,\\
&&${m_{(1,0,2)}^{(1,1)}}^2-{m_{(1,0,2)}^{(3,1)}}^2={m_{(1,1,1)}^{(1,1)}}^2-{m_{(1,1,1)}^{(3,1)}}^2$\\
&&$~~~={m_{(2,1,1)}^{(3,1)}}^2-{m_{(2,1,1)}^{(1,1)}}^2= {m_{(2,0,2)}^{(3,1)}}^2-{m_{(2,0,2)}^{(1,1)}}^2$\\
\cline{2-3}
&(3,2,2,2)&$\displaystyle{\sum_{\alpha=1}^2{m_{(2,1,1)}^{(\alpha,1)}}^2-\sum_{\beta=1}^2{m_{(1,2,0)}^{(1,\beta)}}
=\sum_{\beta=1}^2{m_{(1,0,2)}^{(1,\beta)}}^2-\sum_{\alpha=1}^2{m_{(0,1,1)}^{(\alpha,1)}}^2}$\\
&&$~~~={m_{(2,0,0)}^{(1,1)}}^2-{m_{(2,0,2)}^{(1,1)}}^2$,\\
&&$\displaystyle{\sum_{\alpha=1}^2{m_{(2,1,1)}^{(\alpha,1)}}^2+\sum_{\alpha=1}^2{m_{(0,1,1)}^{(\alpha,1)}}^2
=\sum_{\beta=1}^2{m_{(1,0,2)}^{(1,\beta)}}^2+\sum_{\beta=1}^2{m_{(1,2,0)}^{(1,\beta)}}^2}$,\\
&&$\displaystyle{\sum_{\alpha=1}^2\sum_{\beta=1}^2{m_{(1,1,1)}^{(\alpha,\beta)}}^2=2{m_{(2,0,2)}^{(1,1)}}^2+2{m_{(2,0,0)}^{(1,1)}}^2}$,\\
&&${m_{(1,2,0)}^{(1,1)}}^2-{m_{(1,2,0)}^{(1,2)}}^2={m_{(1,0,2)}^{(1,1)}}^2-{m_{(1,0,2)}^{(1,2)}}^2$\\
&&$~~~={m_{(1,1,1)}^{(1,1)}}^2-{m_{(1,1,1)}^{(1,2)}}^2 ={m_{(1,1,1)}^{(2,1)}}^2-{m_{(1,1,1)}^{(2,2)}}^2$,\\
&&${m_{(2,1,1)}^{(1,1)}}^2-{m_{(2,1,1)}^{(2,1)}}^2={m_{(0,1,1)}^{(1,1)}}^2-{m_{(0,1,1)}^{(2,1)}}^2$\\
&&$~~~={m_{(1,1,1)}^{(1,1)}}^2-{m_{(1,1,1)}^{(2,1)}}^2 ={m_{(1,1,1)}^{(1,2)}}^2-{m_{(1,1,1)}^{(2,2)}}^2$\\
\hline
\end{tabular}
\end{center}
\end{table}
We have classified mass relations into three types, but the form of mass relations is not unique.
For example, we derive the second type relation such as 
$\displaystyle{\sum_{\alpha=1}^3 {m_{(2,0,2)}^{(\alpha,1)}}^2 = \sum_{\alpha=1}^3 {m_{(2,1,1)}^{(\alpha,1)}}^2}$ 
and two third type relations 
$\displaystyle{{m_{(2,0,2)}^{(1,1)}}^2 = {m_{(2,1,1)}^{(3,1)}}^2}$ and
$\displaystyle{{m_{(2,0,2)}^{(2,1)}}^2 = {m_{(2,1,1)}^{(2,1)}}^2}$ from $[8,4]$ for $(3,2,3,0)$.
By using them, three third type relations are written down in Table \ref{t5}.
Mass relations derived from $[9,4]$ for $(p,q,r,s) = (3,2,1,3)$ are obtained from those for $(p,q,r,s) = (3,2,3,1)$ 
by the following replacement,
\begin{eqnarray}
&~& {m_{(1,0,2)}^{(\alpha,1)}}^2 \to {m_{(1,2,0)}^{(1,\beta)}}^2 ,~ {m_{(0,1,3)}^{(1,1)}}^2 \to {m_{(2,1,1)}^{(1,1)}}^2 ,~
{m_{(2,0,2)}^{(\alpha,1)}}^2 \to {m_{(1,1,1)}^{(1,\beta)}}^2 ,
\nonumber \\
&~& {m_{(1,1,1)}^{(\alpha,1)}}^2 \to {m_{(2,0,0)}^{(1,\beta)}}^2 ,~ {m_{(1,2,0)}^{(1,1)}}^2 \to {m_{(1,0,0)}^{(1,1)}}^2 ,~
{m_{(2,1,1)}^{(\alpha,1)}}^2 \to {m_{(0,1,1)}^{(1,\beta)}}^2 ,
\nonumber \\
&~& {m_{(2,2,0)}^{(1,1)}}^2 \to {m_{(2,2,0)}^{(1,1)}}^2 .
\label{rep1}
\end{eqnarray}
In the same way, mass relations derived from $[9,4]$ for $(p,q,r,s) = (3,2,0,4)$ are obtained from those for $(p,q,r,s) = (3,2,4,0)$ 
by the following replacement,
\begin{eqnarray}
&~& {m_{(2,0,2)}^{(\alpha,1)}}^2 \to {m_{(2,0,0)}^{(1,\beta)}}^2 ,~ {m_{(0,1,3)}^{(\alpha,1)}}^2 \to {m_{(1,0,0)}^{(1,\beta)}}^2 ,
\nonumber \\
&~& {m_{(2,1,1)}^{(\alpha,1)}}^2 \to {m_{(1,2,0)}^{(1,\beta)}}^2 ,~ {m_{(2,2,0)}^{(1,1)}}^2 \to {m_{(2,2,0)}^{(1,1)}}^2 .
\label{rep2}
\end{eqnarray}
We have obtained mass relations among sfermions which stem from the bulk field with $[N,k]$.
Those are specific to each $[N,k]$ and gauge symmetry breaking pattern, and can be useful probes to select models.

Brane fields at $y=0$ are $SU(5) \times SU(N-5)$ multiplets, and those soft masses satisfy the $SU(5)$ GUT relations,
\begin{eqnarray}
 {m_{\tilde{q}_L}}^2 = {m_{\tilde{u}_R^*}}^2 = {m_{\tilde{e}_R^*}}^2 ,~~~ {m_{\tilde{l}_L}}^2 = {m_{\tilde{d}_R^*}}^2 .
\label{GUTrelation}
\end{eqnarray}

So far we assume that all zero modes survive after the breakdown of extra gauge symmetries.
In case that particle mixing and/or decoupling occurs, some relations should be modified.
We need further model-dependent analyses to derive specific relations in such a case.

\section{Conclusions}

We have studied sfermion masses from a general framework, based on orbifold family unification models 
under some assumptions regarding the breakdown of SUSY and gauge symmetries, and derived relations among them.
Sfermion mass relations are specific to each model and can be useful for a selection of realistic model.

A non-abelian subgroup such as $SU(r) \times SU(s)$ of $SU(N)$ plays the role of family symmetry
and its $D$-term contributions spoil the mass degeneracy.
Conversely, the requirement of degenerate masses would give a constraint on the $D$-term condensations
and/or SUSY breaking mechanism.
For example, if we take Scherk-Schwarz mechanism for $N=1$ SUSY breaking, the $D$-term condensations
vanish for the gauge symmetries broken at the orbifold breaking scale $M_U$
because of a universal structure of soft SUSY breaking parameters.
If extra gauge symmetries, however, are broken at different scales from $M_U$, 
soft SUSY breaking parameters receive extra renormalization effects and turn out a non-universal structure.
As a result, $D$-term contributions can appear.
In this case, our analysis should be modified by considering the renormalization group running for sfermion masses.
In the case that effects such as $F$-term contributions and/or higher dimensional operators are sizable, 
we should consider them.

Sum rules of sparticle masses at the TeV scale can be derived if the physics 
between the breaking scale $M_U$ and the weak scale is specified.
In our previous analysis, we have assumed the gravity-mediated SUSY breaking 
in the case that the dynamics in the hidden sector do not give sizable effects
on renormalization group evolutions of soft SUSY breaking parameters.\cite{K&Ki}
It is also important to study the case with strong dynamics in the hidden sector.\cite{CR&S}

\section*{Acknowledgements}
This work was supported in part by Scientific Grants from the Ministry of Education and Science, 
Grant No.18204024, Grant No.18540259 (Y.K.).

\end{document}